\documentclass{article}
\title{Optimizing Compiler for Engineering Problems}
\author{Petr R. Ivankov}
\date{August 22, 2008}

\usepackage{amsmath,amsthm, pb-diagram, lamsarrow, pb-lams, hyperref}
\usepackage{graphicx}
\usepackage{tabularx}
\usepackage{float}

\input{amssymb.sty}
\chardef\bslash=`\\ 

\theoremstyle{definition}

\theoremstyle{remark}

\def\7{\dagger}

\newcommand{\eval}[2][\right]{\relax
  \ifx#1\right\relax \left.\fi#2#1\rvert}

\begin{document}
\maketitle
\markboth{Petr R. Ivankov}{Runtime Optimizing Compiler for Engineering Problems}

\renewcommand{\sectionmark}[1]{}

\begin{abstract}
{
New information technologies provide a lot of prospects for performance improvement.
One of them is "Dynamic Source Code Generation and Compilation". This article shows
how this way provides high performance for engineering problems.
}
\end{abstract}

\section{Introduction}

Present day engineering problems require multidisciplinary usage of computer. For example engineers use following IT branches:
math calculations, 3D graphics, real time control, databases. Software devoted for every single branch is not effective.
This circumstance reflects common integration strategy of IT. There exists a myth that performance requires specialization and contradicts universality and/or integration strategy.
This article shows that very universal software can provide very high performance.
Section 2 is devoted to universal software. Section 3 describes runtime optimizing for it.

\section{Universal Framework for Science and Engineering}

Information about universal engineering software is contained at following web cites.
\begin{itemize}
\item Universal Framework for Science and Engineering homepage \break
 \url{http://www.mathframe.com/};
\item CodeProject articles devoted to framework applications \break  \url{http://www.codeproject.com/script/articles/list_articles.asp?userid=3080377};
\item Arxiv articles \cite{maggregates}, \cite{vrf} devoted to framework applications.
\item SourceForge projects \break
AstroFrame \url{https://sourceforge.net/projects/astrohalaxy};
\break
CategoryTheory \url{https://sourceforge.net/projects/categorytheory/};

\end{itemize}

Described framework is based on three principles. First one is component approach. Second principle is insertion of math formulas.
Third principle is openness
of framework. So let us consider them.

\subsection{Component approach}

The best method of complicated phenomenon grasping is a decomposition of the phenomenon. The phenomenon contains objects (components)
and there exist links between objects. Any object may belong to a set of domains. For example a source of physical field
\cite{field} has a geometrical position. Hence it is a subject of positioning domain.
This object may be linked to other object of positioning domain by
positioning links. If source of physical field receives and then transmits information then it is a consumer of information.
So it is also a subject
of information domain and may be linked to sources of information. And at last it is a subject of physical field domain.
Typical picture of objects (components) of virtual reality simulation is presented on Figure 1.
\begin{figure}[htb]
\begin{center}
 \fbox{\includegraphics [scale=0.5]{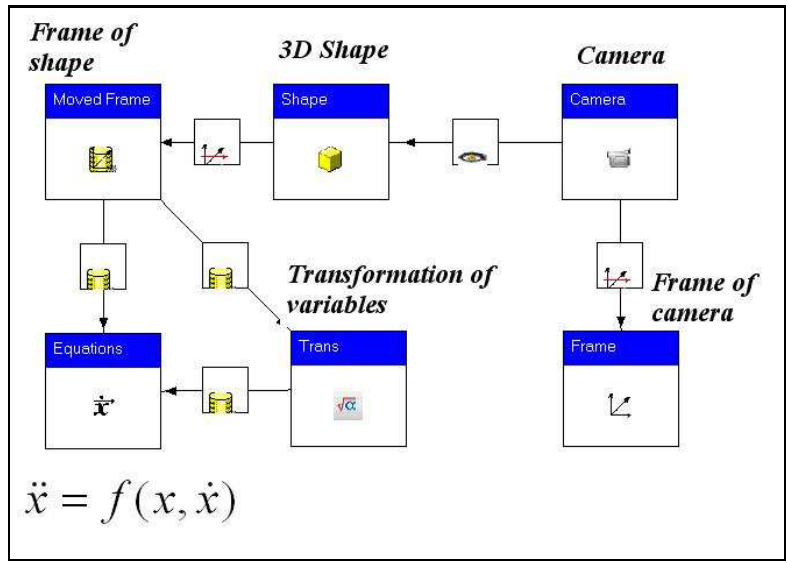}}
 \caption {Typical example of virtual reality}
\end{center}
\end{figure}
This picture shows motion of 3D shape. We have ordinary differential equations of the shape motion

\begin{equation}
\ddot{x}=f(x, \dot{x}). \nonumber \
\end{equation}

The component at left bottom is a solver of these equations. It is a source of information. In this situation
we should perform some transformation of this information. To do this we use ``Transformation of variables".
Later component is a consumer of information of the solver. So it is connected to solver by information link.
We also have a ``Frame of shape". It is a moved reference frame that uses information form the solver and from the
``Transformation of variables". So it is a consumer of information and it is connected to its providers of information.
The ``3D Shape" is rigidly connected to this frame. The shape is connected to the frame by geometrical positioning link.
Also we have a virtual camera and its reference frame. They are also connected by geometrical positioning link. And at
last the camera is connected to the shape by visibility link.

\subsection{Formula editor}

We need rich formula editor for interdisciplinary software.
Insertion of formulas enables us use them in different tasks of virtual reality.
These formulas may be used for definition of signals, transformation of 3D figures, definition of right parts of differential equations,
and definition of size and color of figures, etc.
The formulas may contain real, integer, boolean, vector variables etc. In the CategoryTheory project \url{https://sourceforge.net/projects/categorytheory/} formula,
the editor operates even with Galois fields.
Implemented in related projects formula editor formula editor is case sensitive and operates with a lots of types.
Result of calculation depends on types of variables.
Let $\sin (a)$ be a formula of formula editor. What is its meaning?
If $a$ is a real variable then result of formula is a real value.
However if $a$ is an array then $\sin (a)$ is also an array of componentwise calculation of $\sin$.
If $a$ is a real array and $b$ is a real variable then $a + b$ means an array of sums of components of $a$ with $b$.
Any function of formula editor may be a variable or result of calculation.
A function as result is not a value of function but it is the whole function.
This fact seems unusual for people who do not know functional analysis.
For example if $a$ is a real variable then $f(a)$ means result of calculation of $f$.
If $a$ is a function then $f(a)$ is a composition of functions.
Formula editor supports matrix and vector operations.
Examples of usage of vector and matrix operations are presented below:
\begin{equation}
f^taf, \nonumber \
\end{equation}
\begin{equation}
(q^{-1}+h)^{-1}, \nonumber \
\end{equation}
\begin{equation}
a\times b. \nonumber
\end{equation}
These examples contains transposition of matrixes, products of matrixes, inversion of matrixes and vector product of 3D vectors.
A very good sample of these operations' applications is Kalman filter \cite{Kalmanfilter}. In particular this filter is used in motion
control systems. You can download and evaluate example of this filter from:
\url{http://www.codeproject.com/cs/library/UniversalEnggFrmwork6.asp}
The formula editor supports Dirac delta function. The presence of the delta function at the right part of the ordinary differential equation shows
that the result function is not continuous.
Following picture shows presence of delta function in formula editor
\begin{equation}
f(t)\delta (t) \nonumber \
\end{equation}

\subsection{Openness of the software}
Usually, every developer or company has its own projects of those object domain.
This software does not require discarding of existing projects. Any object domain project may be included to this software.
If you wish to include your project or its part, then you should develop an adapter, compile the class library, and link it to this software.
The adapter should contain one or more classes that implement one or more interfaces of this software.
A more profound description of these interfaces is contained in the developer's guide. You can download the guide from
AstroFrame homepage \url{https://sourceforge.net/projects/astrohalaxy}.

\section{Optimizing Compiler}

Optimizing compiler transforms formula editor expressions to C\# code. Then CodeDom \break
\url{http://msdn.microsoft.com/en-us/library/650ax5cx.aspx} \break
performs compilation. Generated code do not contain cycles or unnecessary assignments. Let us consider some examples of generated code.

\subsection{Matrix Product}

Formula editor indicates matrix product as $ab$ where $a$ and $b$ are first and second factor respectively.
Dimension of $a$ and $b$ depends on framework project. Generated code looks like: \break
\texttt{var\_2[0, 0] = var\_0[0, 0] * var\_1[0, 0] + var\_0[0, 1] * var\_1[1, 0] + }
\texttt{var\_0[0, 2] * var\_1[2, 0] + var\_0[0, 3] * var\_1[3, 0];} \break
\texttt{var\_2[0, 1] = var\_0[0, 0] * var\_1[0, 1] + var\_0[0, 1] * var\_1[1, 1] + }
\texttt{var\_0[0, 2] * var\_1[2, 1] + var\_0[0, 3] * var\_1[3, 1];} \break
... \break
where \texttt{var\_0, var\_1} and \texttt{var\_2} corresponds to $a$, $b$ and $ab$ respectively
So we have code without cycles and unnecessary assignments. It is possible code generation even without arrays.
\subsection{Vector Calculation}

Formula editor is context sensitive. It has high level of abstraction. For example if $a$ and
$b$ are vectors then $a + b$ is vector sum. But if $a$ is a scalar then $a + b$ means componentwise
sum to scalar. Following code shows this situation:
\texttt{var\_2[0] = var\_0 + var\_1[0];} \break
\texttt{var\_2[1] = var\_0 + var\_1[1];} \break
\texttt{var\_2[2] = var\_0 + var\_1[2];} \break
... \break

where \texttt{var\_0, var\_1} and \texttt{var\_2} corresponds to $a$, $b$ and result of calculation 
respectively.

\subsection{Sum}
If $a$ is number array then formula editor can represent sum of its elements by the following way
\begin{equation}
\sum a;
\end{equation}
\texttt{var\_1 = var\_0[0] + var\_0[1] + var\_0[2] + var\_0[3] + ...;}\break
where \texttt{var\_0, var\_1} corresponds to $a$ and $\sum a$ respectively.
Generated code is presented below:

\subsection{Symbolic Differentiation}

Framework contains symbolic differentiation.
Expression $\frac{d}{dt}\sin t^2$ match to the following code.
\texttt{var\_1 = 2 * var\_0 * Math.Sin(var\_0);} \break
where \texttt{var\_0, var\_1} corresponds to $t$ and result of differentiation.

\subsection{Conclusion}

There exists opinion that .NET does not match to scientific and engineering problems.
One reason that .NET requires garbage collection. However the framework construct new objects
during project loading only. Main calculations do not contain construction and therefore they do
not require garbage collection. Second reason is that .NET has slow performance. This article shows
that runtime compilation enables us reach very high performance.
Besides this circumstances .NET have a lot of other properties that are useful for
scientific and engineering problems. Thank you very much.

\end{document}